\documentclass[aps,prl,a4paper,reprint,amsmath,nofootinbib,longbibliography,twocolumn]{revtex4}

\usepackage{calc}
\usepackage{amsmath}
\usepackage{graphicx}
\usepackage{makecell}
\usepackage{bm}
\usepackage[normalem]{ulem}
\usepackage{booktabs}
\usepackage[inline]{enumitem}
\usepackage[hidelinks]{hyperref}

\let\oldbibitem\bibitem
\renewcommand{\bibitem}{%
  \renewcommand{\doi}[1]{doi \href{https://doi.org/##1}{##1}}
  \let\bibitem\oldbibitem
  \oldbibitem
}

\renewcommand{\vec}[1]{\ensuremath{\bm{#1}}}
\newcommand{\mat}[1]{\ensuremath{\bm{#1}}}
\newcommand{\cmmd}{{CM$^{\text{md}}$ }}
\newcommand{\matern}{Mat{\'e}rn}
\newcommand{\twocol}[1]{\multicolumn{2}{@{}c@{}}{#1}}

\usepackage{xcolor}

\usepackage{etoolbox}

\patchcmd{\section}{\centering}{\raggedright}{}{}
\patchcmd{\section}{0.5cm}{1mm}{}{}
\patchcmd{\subsection}{\centering}{\raggedright\vspace*{-\baselineskip}}{}{}
\patchcmd{\subsection}{.5cm}{1mm}{}{}
\renewcommand{\subsubsection}[1]{\addcontentsline{toc}{subsubsection}{\hspace{5pt}#1}}

\newcommand{\highlight}[1]{\emph{#1}}

\setcitestyle{numbers,square,citesep={,\kern-.24em}}  

\AtBeginDocument{%
  \heavyrulewidth=.08em
  \lightrulewidth=.05em
  \cmidrulewidth=.03em
  \belowrulesep=.65ex
  \belowbottomsep=0pt
  \aboverulesep=.4ex
  \abovetopsep=0pt
  \cmidrulesep=\doublerulesep
  \cmidrulekern=.5em
  \defaultaddspace=.5em
}

\usepackage{pdfpages}

\begin{document}


\title{Unified Representation of Molecules and Crystals for Machine Learning}

\author{Haoyan Huo}
\affiliation{School of Physics, Peking University, Beijing, China\\Present address: Department of Materials Science and Engineering, University of California, Berkeley, USA}

\author{Matthias Rupp}
\affiliation{Fritz Haber Institute of the Max Planck Society, Berlin, Germany\\ Department of Computer and Information Science, University of Konstanz, Germany\\ Present address: Materials Research and Technology Department, Luxembourg Institute of Science and Technology (LIST), Belvaux, Luxembourg}
\email{mrupp@mrupp.info}
\homepage{www.mrupp.info}

\date{\today}

\begin{abstract}\noindent
Accurate simulations of atomistic systems from first principles are limited by computational cost.
In high-throughput settings, machine learning can reduce these costs significantly by accurately interpolating between reference calculations.
For this, kernel learning approaches crucially require a representation that accommodates arbitrary atomistic systems.
We introduce a many-body tensor representation that is invariant to translations, rotations, and nuclear permutations of same elements, unique, differentiable, can represent molecules and crystals, and is fast to compute.
Empirical evidence for competitive energy and force prediction errors is presented for changes in molecular structure, crystal chemistry, and molecular dynamics using kernel regression and symmetric gradient-domain machine learning as models.
Applicability is demonstrated for phase diagrams of Pt-group/transition-metal binary systems.

\bigskip

This study has been published. See \href{https://doi.org/10.1088/2632-2153/aca005}{DOI 10.1088/2632-2153/aca005} for the version of record.
\end{abstract}

\maketitle

\enlargethispage*{2\baselineskip}


\section{Introduction}

\subsubsection{Rationale: ML boosts QM calculations}\noindent
The computational study of atomistic systems such as molecules and crystals requires accurate treatment of interactions at the atomic and electronic scale.
Accurate first-principles methods, however, are limited by their high computational cost.
In settings that require many calculations, such as dynamics simulations, phase diagrams, or high-throughput searches, machine learning (ML) \cite{g2015,jm2015} can reduce overall costs by orders of magnitude via accurate interpolation between reference calculations. \cite{jkk2019q,scaccr2019q,rdrl2015}
For this, the problem of repeatedly solving a complex equation such as Schr{\"o}dinger's equation \highlight{for many related inputs} is mapped onto a non-linear regression problem:
Instead of numerically solving new systems, they are statistically estimated based on a reference set of known solutions.~\cite{rtml2012,r2015}
This ansatz enables, among other applications, screening larger databases of molecules and materials~\cite{rdrl2014,rdrl2015}, running longer dynamics simulations~\cite{bpkc2010}, investigating larger systems~\cite{rrl2015}, and increasing the accuracy of calculations~\cite{bgmc2013,rdrl2015}.

\subsubsection{Need for representations, and their requirements}
Kernel-based ML models \cite{r2015fq,ctspsm2017,deringer2021q,unke2021q} for data-efficient accurate prediction of {ab initio} properties require a single space in which regression is carried out.
Representations \cite{lgr2021} are functions that map atomistic systems to elements in such spaces, either directly or via a kernel \cite{kernelspace}.
Representations should be
(i)~\emph{invariant} against transformations preserving the predicted property, in particular translations, rotations, and nuclear permutations of same elements,
as learning these invariances from data would require many reference calculations; non-scalar properties can require equivariance instead of invariance;
(ii)~\emph{unique}, that is, variant against transformations changing the property, as
systems with identical representation that differ in property would introduce errors~\cite{m2012e};
(iii)~\emph{continuous}, and ideally \emph{differentiable},
as discontinuities work against the smoothness assumption of the ML model and model gradients are often useful;
(iv)~\emph{general} in the sense of being able to encode any atomistic system, including finite and periodic systems;
(v)~\emph{fast} to compute, as the goal is to reduce computational cost;
(vi)~\emph{data-efficient} in the sense of requiring few reference calculations to reach a given target error.
\ Constant size is an advantage, \cite{cgvy2017} as is the ability to encode the whole system as well as local atomic environments.
Requirements (v) and (vi) are in practice determined empirically.
See Refs.~\citenum{bkc2013,lrrk2015,ook2020q,lgr2021} for details on these and further requirements.

\subsubsection{Representations and descriptors}
Some representations fulfill these requirements only partially,
such as the Coulomb matrix (CM) \cite{rtml2012} and bag of bonds (BoB) \cite{hbrplmt2015} discussed below.
State-of-the-art representations often fulfill these requirements in some limit, such as infinite expansion order.
See Ref.~\citenum{lgr2021} for a comprehensive and detailed discussion.
The descriptors used in cheminformatics, \cite{tc2009} and sometimes in materials informatics, often violate (ii) and (iii), in particular if they do not include atomic coordinate information or rely on cutoff-based definitions of chemical bonds.
Such descriptors serve the different purpose of predicting derived properties that are not functions of a single conformation, such as solubility or binding affinity to a macromolecule.

\subsubsection{Scope: Content of study}
We introduce a many-body tensor representation (MBTR) derived from CM/BoB and concepts of many-body expansions.
It is related \cite{lgr2021} to Behler-Parrinello symmetry functions \cite{bp2007} and histograms of distances, angles, and dihedral angles \cite{fhhgsdvkrv2017}.
MBTR fulfills the above requirements in the limit, is interpretable, allows visualization (Figure~\ref{figVisMbtr}), and describes finite and periodic systems.
State-of-the-art empirical performance is demonstrated by us for organic molecules and inorganic crystals, as well as applicability to phase diagrams of Pt-group\,/\,transition metal binary systems,
and by others for predicting and optimizing various molecular \cite{lumiaro2021q,bahlke2020q,petry2021,lourenco2021q,lourenco2021bq,iype2019q,zhai2020q} and crystalline properties \cite{hxh2020,mayr2021q,arrigoni2021q,phlnmkh2020q}.
Implementations of MBTR are publicly available (see Code Availability section at the end).


\section{Method}

\subsubsection{CM and BoB shortcomings}\noindent
We start from the CM~\cite{rtml2012,mrgvmhtmvl2013,rrl2015}, which represents a molecule~$\mathcal{M}$ as a symmetric atom-by-atom matrix
\begin{equation}\label{equCM}
	\mat{M}_{i,j} = \begin{cases} \frac{1}{2} Z_i^{2.4} & i = j \\[3pt] \frac{Z_i Z_j}{d_{i,j}} & i \neq j \end{cases} ,
\end{equation}
where $Z_i$ are atomic numbers and $d_{i,j} = ||\vec{R_i}-\vec{R_j}||$ is Euclidean distance between atoms $i$ and $j$.
To avoid dependence on atom ordering (in the input), which would violate (i), $\mat{M}$ is either diagonalized, loosing information which violates~(ii)~\cite{m2012e}, or sorted, causing discontinuities that violate~(iii).
Another shortcoming is the use of $Z$, which is not well suited for interpolation \cite{gvlds2015} as it overly decorrelates chemical elements from the same column of the periodic table.

The related BoB \cite{hbrplmt2015} representation uses the same terms, but arranges them differently.
For each pair of chemical elements, corresponding CM terms
are stored in sorted order, 
which can be viewed as an $N_e \times N_e \times d$ tensor, or an $N_e \times (N_e + 1)/2 \times d$ tensor if symmetry is taken into account, 
where $N_e$ is number of elements and $d$ is sufficiently large.
Unlike the CM, it can not distinguish homometric molecules \cite{lrrk2015}, which might distort its feature space \cite{pwbocc2020q}.
%
While the BoB tensor itself does not suffer from discontinuities, its derivative does.
%

\subsubsection{Definition}
To derive MBTR, we retain stratification by elements, but avoid the sorting by arranging distances on a real-space axis:
\begin{equation}
	\label{equBobTrep}
	f_{\text{BoB}}\bigl(x,z_1,z_2\bigr) = \sum_{i,j=1}^{N_a} \delta\bigl(x-d_{i,j}^{\,-1}\bigr) \delta(z_1,Z_{i}) \delta(z_2,Z_{j}) ,
\end{equation}
where $x$ is a real number, $z_1,z_2$ are atomic numbers, $N_a$ is number of atoms, $\delta(\cdot)$ is Dirac's delta, and $\delta(\cdot,\cdot)$ is Kronecker's delta.
$f_{\text{BoB}}$ has mixed continuous-discrete domain and encodes all (inverse) distances between atoms with elements $z_1$ and $z_2$.
For a smoother measure, we replace Dirac's~$\delta$ with another probability distribution~$\mathcal{D}$, ``broadening'' or ``smearing'' it. \cite{bkc2013,csrmrdkl2015}
In this work, we use the normal distribution.
Other distributions can be used, in particular symmetric and short-tailed ones, for example, the Laplace distribution or the uniform distribution.
We did not observe significant differences in performance, however. 
Adding a weighting function~$w_2$ and replacing the Kronecker~$\delta$ functions by an element correlation matrix $C \in \bm{R}^{N_e \times N_e}$ yields
\begin{equation}
	\label{equBobTrepSmeared}
	f_2\bigl(x,z_1,z_2\bigr) \!=\!\! \sum_{i,j=1}^{N_a} \!w_2(i,j) \, \mathcal{D}\bigl(x,g_2(i,j)\bigr) C_{z_1,Z_{i}} C_{z_2,Z_{j}}
\end{equation}
of which \eqref{equBobTrep} is a special case. 
In general, $g_2$ describes a relation between atoms $i$ and $j$, such as their inverse distance, $\mathcal{D}$ broadens the result of $g_2$, and $w_2$ allows to weight down contributions, for example, from far-away atoms.
%
\subsubsection{Many-body expansion}
Building on the idea of many-body expansions, \cite{hl2016,yhp2017} we generalize from $f_2$ in \eqref{equBobTrepSmeared}, which encodes two-body terms, to the MBTR equation
\begin{equation}
	\label{equMBTR}
	f_k(x,\vec{z}) = \sum_{\vec{i}=1}^{N_a} w_k(\vec{i})\mathcal{D}\bigl(x, g_k(\vec{i})\bigr) \prod_{j=1}^k C_{z_{j},Z_{i_j}} ,
\end{equation}
where $\vec{z} \in \bm{N}^k$ are atomic numbers, $\vec{i} = (i_1,\ldots,i_k) \in \{1,\ldots,N_a\}^k$ are index tuples,
and $w_k$, $g_k$ assign a scalar to $k$~atoms in~$\mathcal{M}$.~\cite{scalargeomf}
Canonical choices of $g_k$ for $k \!=\! 1,2,3,4$ are atom counts, (inverse) distances, angles, and dihedral angles.
The element correlation matrices~$C$ allow exploitation of similarities between chemical element species (``alchemical learning''), for example, within the same column of the periodic table. \cite{faber2018,hkyp2019q,christensen2020}

\subsubsection{Distances}
We measure the similarity of two atomistic systems $\mathcal{M}$ and $\mathcal{M'}$ as the Euclidean distance between their representations,
\begin{equation}
	\label{equMBTDistance}
	d_k^{2}(\mathcal{M},\mathcal{M'}) = \sum_{\vec{z}} \int \bigl( f_k(x,\vec{z}) - f_k' (x,\vec{z}) \bigr)^2 \mathrm{d} x.
\end{equation}

\subsubsection{Discretization}
\begin{figure*}
    \begin{minipage}{0.5\linewidth}\centering
        \includegraphics[width=0.9\linewidth]{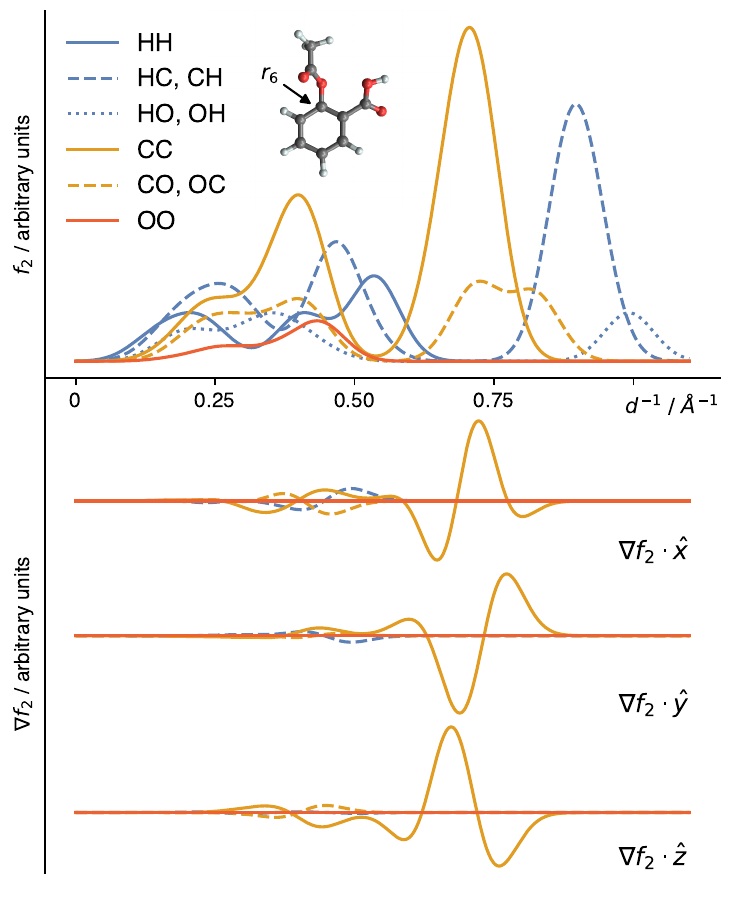}
    \end{minipage}%
    \begin{minipage}{0.5\linewidth}\centering
        \includegraphics[width=0.9\linewidth]{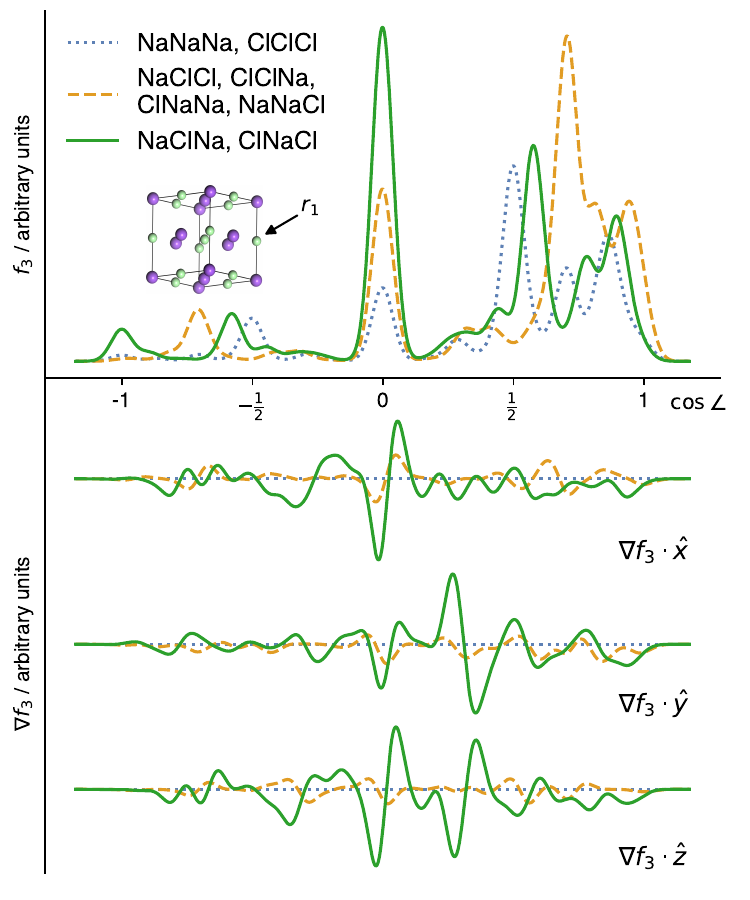}
    \end{minipage}
    \caption{
        \emph{Visualization of many-body tensor representation.}
        Top panels show distributions of inverse distances ($k=2$, quadratic weighting) for aspirin (C$_9$O$_4$H$_8$, left) and distributions of angles ($k=3$, exponential weighting) for fcc salt (NaCl, right).
        Bottom panels show derivatives of the representation, obtained by differentiating with respect to the Cartesian coordinates of C atom~$r_6$ connecting the ester group (left) and the Na atom~$r_1$ at lattice point (right).
    \label{figVisMbtr}}
\end{figure*}

In practice, we adjust \eqref{equMBTR} for symmetries. 
Discretizing the continuous axis as $(x_{\min},x_{\min}+\Delta x,\ldots,x_{\max})$ 
results in a rank $k\!+\!1$ tensor of dimensions $N_e \times \cdots \times N_e \times N_x$ with $N_x = (x_{\max}-x_{\min})/\Delta x$,
where $x_{\min}$ and $x_{\max}$ are the smallest and largest values for which $f_k(x,\vec{z}) \neq 0$ for all $\vec{z}$ and $\mathcal{M}$.
Linearizing element ranks yields $N_e^k \times N_x$ matrices, allowing for visualization (Figure~\ref{figVisMbtr}) and efficient numerical implementation via linear algebra routines.
For systems with many element species, discretization can lead to large matrices, requiring substantial amounts of memory.
In such settings, memory-efficient implementation via sparse matrix formats or on-the-fly calculation of distances and inner products (see, e.g., Ref.~\citenum{faber2018}) of MBTR matrices might be preferable.


\subsubsection{Periodic systems}

Periodic systems, used to model bulk crystals and surfaces, can be viewed as unit cells surrounded by infinitely many translated images of themselves. 
For such systems, $N_a = \infty$ and the sum in \eqref{equMBTR} diverges.
We prevent this by requiring one index of $\vec{i}$ to be in the (same) primitive unit cell.~\cite{indexincell}
This accounts for translational symmetry and prevents double-counting.
Use of weighting functions $w_k$ such as exponentially decaying weights~\cite{expweighting} then ensures convergence of the sum.
Figure~\ref{figVisMbtr}~(right) presents the resulting distribution of angles for face-centered cubic (fcc) NaCl as an example.
Note that the $k$-body terms $g_k$ do not depend on choice of unit cell geometry (lattice vectors).
This ensures unique representation of Bravais lattices where the choice of basis vectors is not unique, for example 2D hexagonal lattices where the angle between lattice vectors can be $\frac{1}{3}\pi$ or $\frac{2}{3}\pi$.

\subsubsection{Derivatives}

Many applications, including dynamics simulations and structural relaxation, require forces, the negative gradient of the energy with respect to atomic coordinates.
The gradient of \eqref{equMBTR} is given by
\begin{multline}
	\label{equMBTRDiv}
	\nabla f_k(x,\vec{z}) 
	= \sum_{\vec{i}=1}^{N_a} \Biggl( \mathcal{D}\bigl(x, g_k(\vec{i})\bigr) \nabla w_k(\vec{i}) \\
	+ w_k(\vec{i}) \frac{\partial\,\mathcal{D}\bigl(x, g_k\bigr)}{\partial g_k } \nabla g_k(\vec{i}) \Biggr) \prod_{j=1}^k C_{z_{j},Z_{i_j}}.
\end{multline}
The gradient $\nabla f_k(x,\vec{z}$) can be derived analytically if this is possible for $\nabla w_k$, $\nabla g_k$, and $\nabla \mathcal{D}$.
Alternatively, automatic differentiation \cite{jax,pytorch} can be used, removing the need for manual derivation. 
Figure~\ref{figVisMbtr} visualizes MBTR gradients.


\section{Results}

\noindent
To validate MBTR, we demonstrate accurate predictions for properties of molecules and crystals.
Focusing on the representation, we employ plain kernel ridge regression models~\cite{r2015} unless stated otherwise.

\subsection{Changes in molecular structure}

To demonstrate interpolation across {changes in the chemical structure of molecules} we utilize a benchmark dataset~\cite{mrgvmhtmvl2013} of 7,211 small organic molecules composed of up to seven C, N, O, S and Cl atoms, saturated with~H.
Molecules were relaxed to their ground state using the Perdew-Burke-Ernzerhof (PBE) \cite{pbe1996} approximation to Kohn-Sham density functional theory (DFT).
Restriction to relaxed structures projects out spatial variability and allows focusing on changes in chemical structure.
Table~\ref{tabOrganicMolecules} presents prediction errors for atomization energies and isotropic polarizabilities obtained from single point calculations with the hybrid PBE0~\cite{peb1996,ab1999} functional.
For 5\,k training samples, prediction errors are below 1\,kcal/mol (``chemical accuracy''), with the MBTR model's mean absolute error of 0.6\,kcal/mol corresponding to thermal fluctuations at room temperature. 
Note that MBTR achieves similar performance with a linear regression model, allowing constant-time predictions.


\begin{table}[b]
    \caption{\emph{Prediction errors for small organic molecules.}
        Machine-learning models of atomization energies~$E$ and isotropic polarizabilities~$\alpha$,
        obtained at hybrid density functional level of theory, were trained on 5\,k molecules and evaluated on 2\,k others using different representations.
        {\scriptsize\selectfont 
        RMSE \!=\! root mean square error, 
        MAE \!=\! mean absolute error, 
        CM \!=\! Coulomb matrix,
        BoB \!=\! bag of bonds, 
        BAML \!=\! bonding angular machine learning, 
        SOAP \!=\! smooth overlap of atomic positions, 
        FCHL19 \!=\! Faber-Christensen-Huang-Lilienfeld representation,
        MBTR \!=\! many-body tensor representation.
        }\label{tabOrganicMolecules}} \bigskip

    \begin{tabular}{llcccccc}
    	\toprule
                               &           && \multicolumn{2}{c}{E / kcal mol$^{-1}$} & & \multicolumn{2}{c}{$\alpha$ / \AA$^3$} \\ 
		\cmidrule(lr){4-5} \cmidrule(lr){7-8} 
        Representation         & Kernel    &&    MAE  &    RMSE  &&    MAE  &    RMSE \\
        \midrule
        CM~\cite{rtml2012}     & Laplacian &&    3.47 &    4.76  &&    0.13 &    0.17 \\
        BoB~\cite{hbrplmt2015} & Laplacian &&    1.78 &    2.86  &&    0.09 &    0.12 \\
        BAML~\cite{hl2016}     & Laplacian &&    1.15 &    2.54  &&    0.07 &    0.12 \\ 
        SOAP~\cite{dbcc2016}   & REMatch   &&    0.92 &    1.61  &&    0.05 &    0.07 \\ 
        FCHL19~\cite{faber2018,christensen2020} & Gaussian && 0.44 & --- && --- & --- \\[1ex] 
        MBTR                   & Linear    &&    0.74 &    1.14  &&    0.07 &    0.10 \\ 
        MBTR                   & Gaussian  &&    0.60 &    0.97  &&    0.04 &    0.06 \\
        \bottomrule
    \end{tabular}
\end{table}

\subsection{Changes in crystal chemistry}

\begin{figure}
    \includegraphics[width=\linewidth,trim={7pt 9pt 7pt 9pt},clip]{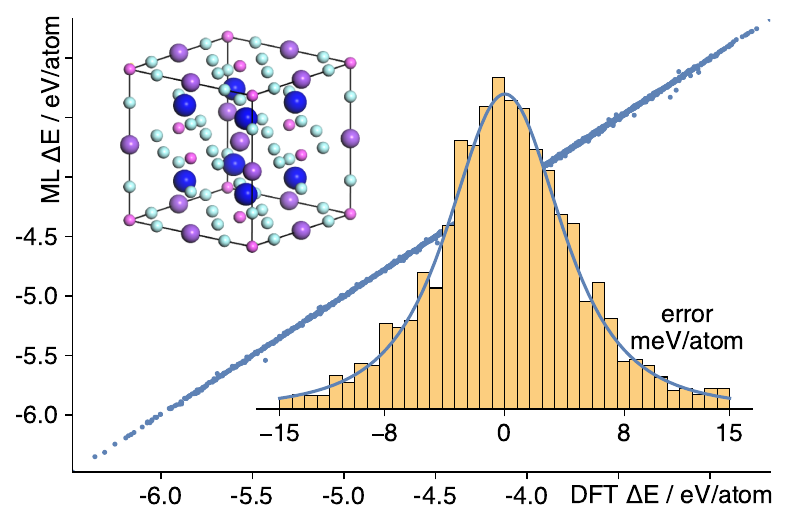}%
    \caption{\emph{Formation energy predictions for ABC$_2$D$_6$ elpasolite structures} containing 12 different elements.
    Shown are reference energies (DFT~$\Delta E$) and predicted energies (ML~$\Delta E$), as well as distribution of errors (inset) for 2\,272 crystals, from an MBTR machine learning model trained on 9\,086 other ones.
    \label{figElpasolite}}
    \vspace*{-3.5mm}
\end{figure}

Interpolation across {changes in the chemistry of crystalline solids} is demonstrated for a dataset of 11\,k elpasolite structures (ABC$_2$D$_6$, AlNaK$_2$F$_6$ prototype) \cite{flva2016,elpasolitedataset} composed of 12 different elements, with geometries and energies computed at DFT/PBE level of theory.
Predicting formation energies with MBTR yields a root-mean-squared-error (RMSE) of 8.1\,meV/atom and mean absolute error (MAE) of 4.7\,meV/atom (Figure~\ref{figElpasolite}) for a training set of 9\,k crystals.

Adding chemical elements should increase the intrinsic dimensionality of the learning problem, and thus prediction errors.
To verify this, we created a dataset of 4\,611 ABC$_2$ ternary alloys containing 22 non-radioactive elements from groups 1, 2, 13--15, spanning five rows and columns of the periodic table.
Structures were taken from the Open Quantum Materials Database (OQMD)~\cite{skamw2013,ksmtdarw2015}, with geometries and properties also computed via DFT/PBE.
As expected, energy predictions exhibit larger errors (RMSE 31\,meV/atom, MAE 23\,meV/atom) compared to an elpasolite model of same training set size (RMSE 23\,meV/atom, MAE 15\,meV/atom).

\subsection{Changes in molecular geometry}

\newcommand{\tmpa}{\hphantom{$^a$}0.07$^a$}
\newcommand{\tmpb}{\hphantom{$^a$}0.06$^a$}

\begin{table*}
    \caption{\emph{Energy and force prediction errors for changes in geometry of organic molecules.}
        Shown are prediction errors for total energies (kcal/mol) and atomic forces (kcal/mol/\AA).
        {\scriptsize\selectfont
        \quad
        MAE  \!=\! mean absolute error, 
        RMSE \!=\! root mean squared error,
        PaiNN \!=\! polarizable atom interaction neural network \cite{sug2021q},
        FCHL19 \!=\! Faber-Christensen-Huang-Lilienfeld representation \cite{faber2018,christensen2020},
        sGDML \!=\! symmetric gradient domain machine learning \cite{ctspsm2017},
        s\matern \!=\! \matern{} kernel augmented with symmetric permutations for sGDML \cite{chmiela2018},
        CM$^{\text{md}}$ \!=\! Coulomb matrix variant, 
        MBTR \!=\! many-body tensor representation.}
    \label{tabGeometryOrganicMolecules}}

    \medskip

    \begin{minipage}{\linewidth}
	    \begin{tabular}{@{} l @{\hspace{6\tabcolsep}} cc @{\hspace{6\tabcolsep}} cc @{\hspace{6\tabcolsep}} cc @{\hspace{6\tabcolsep}} cc @{\hspace{6\tabcolsep}} p{1em} @{\hspace{6\tabcolsep}} c @{\hspace{4\tabcolsep}} c @{\hspace{4\tabcolsep}} c @{}}
	    	\toprule
	        & \multicolumn{8}{@{}c@{}}{Trained only on forces, 1\,k refs.} &
	        & \multicolumn{3}{@{}c@{}}{Trained only on energies, 10\,k refs.} 
	        \\ \cmidrule(lr){2-9} \cmidrule(lr){11-13}

			& \twocol{PaiNN}
	        & \twocol{FCHL19}
	        & \twocol{sGDML/\cmmd}
	        & \twocol{sGDML/MBTR}
	        &
	        & {\cmmd} 
	        & {MBTR} 
	        & {MBTR} \\

	        Kernel
	        & \twocol{---}
	        & \twocol{Gaussian}
	        & \twocol{s\matern}
	        & \twocol{s\matern}
	        &
	        & {Gaussian} 
	        & {linear}    
	        & {Gaussian}     
	        \\ 
	        \cmidrule(lr){2-3} \cmidrule(lr){4-5} \cmidrule(lr){6-7} \cmidrule{8-9} \cmidrule(lr){11-11} \cmidrule(lr){12-12} \cmidrule(lr){13-13}

			               & Energy  & Force   & Energy  & Force   & Energy   & Force  &  Energy  & Force   && Energy & Energy   & Energy  \\
			Molecule       & MAE     & MAE     & MAE     & MAE     & MAE      & MAE    &  MAE     & MAE     && MAE    & MAE      & MAE     \\ \midrule
			benzene        & ---     & ---     & ---     & ---     & \tmpa    &  \tmpb &     0.07 &    0.15 && 0.03   &    0.03  &    0.03 \\
			uracil         &    0.11 &    0.13 &    0.10 &    0.10 &    0.11  &  0.24  &     0.11 &    0.17 && 0.05   &    0.10  &    0.03 \\
			naphthalene    &    0.12 &    0.08 &    0.12 &    0.15 &    0.12  &  0.11  &     0.11 &    0.09 && 0.12   &    0.10  &    0.07 \\
			aspirin        &    0.17 &    0.34 &    0.17 &    0.50 &    0.19  &  0.68  &     0.17 &    0.48 && 0.36   &    0.21  &    0.25 \\
			salicylic acid &    0.12 &    0.20 &    0.12 &    0.22 &    0.12  &  0.28  &     0.11 &    0.18 && 0.11   &    0.13  &    0.07 \\
			malonaldehyde  &    0.10 &    0.34 &    0.08 &    0.25 &    0.10  &  0.41  &     0.09 &    0.36 && 0.18   &    0.21  &    0.10 \\
			ethanol        &    0.06 &    0.22 &    0.05 &    0.14 &    0.07  &  0.33  &     0.06 &    0.26 && 0.17   &    0.17  &    0.06 \\
			toluene        &    0.10 &    0.09 &    0.10 &    0.20 &    0.10  &  0.14  &     0.09 &    0.13 && 0.16   &    0.11  &    0.10 \\      
       
			\bottomrule
	    \end{tabular}
    \end{minipage}

    \begin{flushleft}
    \footnotesize{
	    $^a$ We observed higher noise in predictions of benzene, whose reported prediction errors are also inconsistent in different publications.
    	To make results comparable, we retrained the sGDML/\cmmd model (originally reported values are 0.10 and 0.06).
    }
    \end{flushleft}
\end{table*}

\begin{figure*}[t]
	\includegraphics[width=\linewidth,trim={0pt 0pt 0pt 0pt},clip]{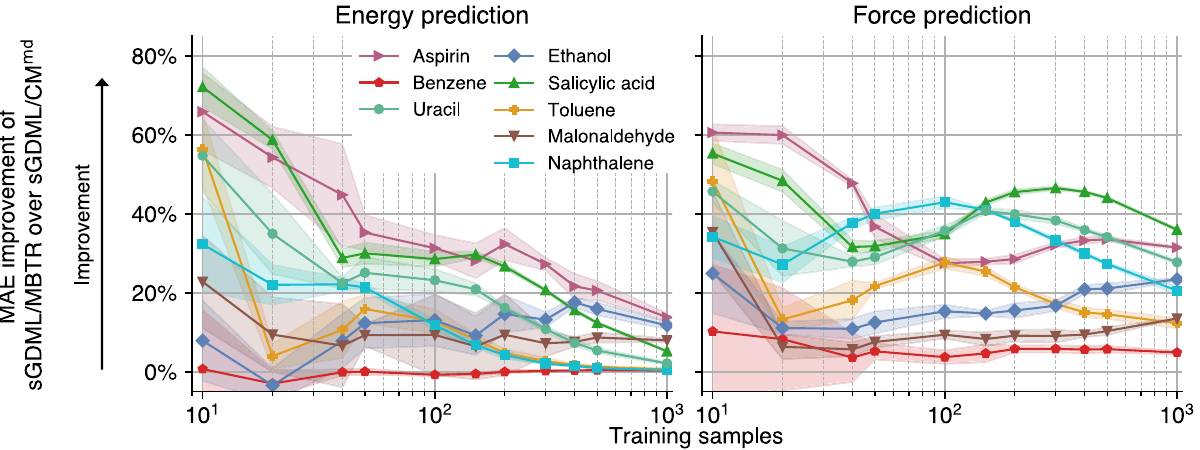}
    \caption{\emph{Relative improvement in predictive accuracy on dynamics data of eight different organic molecules.}
    Shown are force and energy prediction MAE ratios of sGDML/MBTR over sGDML/{\cmmd} as a function of training set size.
    Error bars show the standard deviation of the ratios over five runs with different random seeds.
    \label{figMBTRsGDML}}
    \vspace*{-3.5mm}
\end{figure*}

For interpolation of {changes in molecular geometry}, we employ a benchmark dataset~\cite{sacmt2017,ctspsm2017} of {ab initio} molecular dynamics trajectories of eight organic molecules.
Each molecule was simulated at a temperature of 500\,K for between 150\,k to 1\,M time steps of 0.5\,fs, with energies and forces computed at the DFT/PBE level of theory and the Tkatchenko--Scheffler model~\cite{ts2009} for van der Waals interactions.
Table~\ref{tabGeometryOrganicMolecules} presents results for models trained and evaluated only on energies (right-hand side) and only on forces (left-hand side).

Energy-only models were trained on 10\,k configurations and validated on 2\,k other ones,
employing MBTR (parametrized for dynamics data, see supplement) and a similarly modified CM (\cmmd, see supplement).
Non-linear MBTR regression performs best overall, with the linear kernel again being competitive.

On the one hand, differentiating energy-based machine learning potentials can introduce errors, 
for example, from small oscillations between training samples due to insufficient regularization, 
and from insufficient model constraints in directions not covered by the training data \cite{srhmb2012}.
On the other hand, electronic structure calculations often provide reference forces at not additional cost.
It is therefore beneficial to include these in model training.
This often reduces the required number of reference calculations by an order of magnitude. \cite{bpkc2010,ctspsm2017,chmiela2018,gsd2017}

Force-only models require an adaptation of plain KRR.
To accommodate forces into training and to demonstrate use of MBTR with other regression approaches,
we employ MBTR in the framework of symmetrized gradient-domain machine-learning (sGDML) \cite{chmiela2018}.
This approach uses the {\matern} kernel, augmentation by symmetric molecular permutations, and reference forces for training, while providing energy and force predictions.

Table~\ref{tabGeometryOrganicMolecules} (left-hand side) compares performance of the original sGDML approach (based on the \cmmd{} representation \cite{chmiela2018,ctspsm2017}) and of sGDML based on a 2-body MBTR representation.
Both models were trained on 1000 configurations, leading to kernel matrices with dimensionalities between 27\,k and 63\,k.
For reference, we also present results for the Faber-Christensen-Huang-Lilienfeld (FCHL) representation \cite{faber2018,christensen2020} and the polarizable atom interaction neural network (PaiNN) \cite{sug2021q}.

sGDML/MBTR performs as good or better than sGDML/{\cmmd} for energy and force predictions.
Compared to PaiNN, sGDML/MBTR performs better for energy predictions, but worse for forces.
For a more fine-grained comparison between the sGDML models, Figure~\ref{figMBTRsGDML} presents learning curves of relative MAE ratios of sGDML/MBTR over sGDML/{\cmmd}, together with standard deviations over five runs starting from different random seeds.
sGDML/MBTR consistently outperforms sGDML/{\cmmd} with error reductions up to 50\%--60\%, especially when less than 100 training samples are used.
For more symmetric molecules such as benzene, malonaldehyde, and ethanol, use of MBTR is less beneficial but still an improvement.

\subsection{Phase diagrams}

We demonstrate {applicability} by identifying the convex hull of the phase diagram for Pt-group/transition metal binary alloys, relevant for industrial applications \cite{hcml2013}.
For a given dataset of candidate structures, we predict the energy of each structure and identify those with the lowest energy, which form the convex hull in a phase diagram.
Compositions that lie on or slightly above the convex hull correspond to stable and meta-stable alloys, respectively.

To demonstrate this, we use a dataset \cite{hcml2013} of 153 alloys computed at the DFT/PBE level of theory. 
This dataset contains at most a few hundred structures for each alloy.
Due to this small amount of data direct application of ML models results in errors in predicted energies that are large enough to lead to wrong convex hulls.
We address this by employing a simple active learning \cite{s2012b} scheme:

Starting with a few randomly selected structures, we iteratively train ML models on these and predict energies of candidate structures. 
In each iteration, we calculate (look up) DFT energies only for structures predicted to be low in energy and include these in the training dataset of the next iteration. 
This procedure prevents computationally expensive DFT calculations for high-energy structures that lie above the convex hull, saving up to 48\,\% of all DFT calculations while still identifying the correct convex hull.

\begin{figure}
	\begin{minipage}{\linewidth}\centering
		\includegraphics[width=0.95\linewidth,trim={3pt 0pt 40pt 20pt},clip]{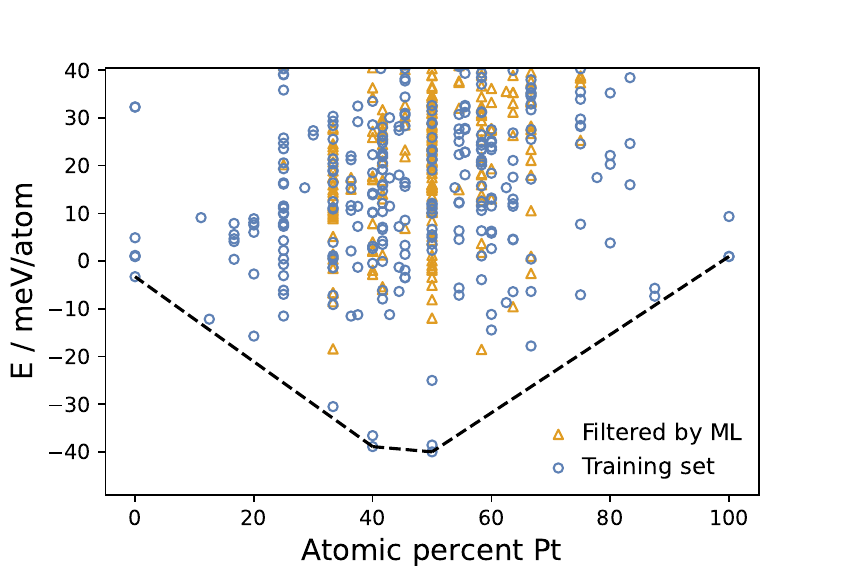}
	\end{minipage}
	\caption{\emph{Phase diagrams of Pt-group/transition metal binary alloys.}
		Shown are Ag$_x$Pt$_{1-x}$ structures (points) and their convex hull (dashed line) as given by DFT and identified by ML.
		All structures are shown at their DFT energy.
		Symbols indicate whether a structure was selected for training (blue circles) or predicted as high-energy (orange triangles).
		See main text for details.
		\label{figPhaseDiagrams}}
\end{figure}

Figure~\ref{figPhaseDiagrams} presents results for AgPt.
The active learning model selected 357 DFT calculations for training and predicted energies of 331 (48\,\%) other structures, with a MAE of 39\,meV/atom.
The trade-off between the number of saved calculations and the probability of failing to identify the correct convex hull can be explicitly controlled by adjusting the energy threshold below which DFT calculations are requested.
In this simple demonstration, structures are given and not derived from composition by relaxation.
While structural relaxation is possible with ML, it brings its own challenges. \cite{ustk2016q,kph2018q,dk2018q,sc2018q,yu2020q,mmbch2020q,huang2022,hao2022,born2021q}

\subsection{Other uses}

\noindent
MBTR has been used to study structure and properties of molecules, clusters, crystals and other atomistic systems.
Studies related to properties include predictions of
\begin{itemize}[wide, labelwidth=!, labelindent=0pt, noitemsep]
	
	\item gas-particle partition coefficients, such as saturation vapour pressure and equilibrium partitioning coefficients, of atmospheric molecules via kernel ridge regression \cite{lumiaro2021q}
	
	\item Heisenberg exchange spin coupling constants for dicopper complexes via Gaussian process regression \cite{bahlke2020q}

	\item total and orbital energies of diverse larger organic molecules from the OE62 dataset \cite{skgtmrro2020q} via a graph neural network \cite{rahaman2020q}
	
	\item extrapolation of size-extensive properties \cite{jskohrm2020q} at the example of atomization energies of organic molecules in the QM9 \cite{rdrl2014} and OE62 \cite{skgtmrro2020q} datasets

	\item formation energies of Al-Ni and Cd-Te binary compounds via support vector regression \cite{hxh2020}

	\item band gaps and formation energies of perovskite-like materials \cite{mayr2021q}

	\item energetics of compositionally disordered compounds via kernel ridge regression \cite{yaghoobi2022}

\end{itemize}

\noindent
Studies related to structure include
\begin{itemize}[wide, labelwidth=!, labelindent=0pt, noitemsep]

	\item visualization of the conformational space of tannic acid molecules via principal component analysis \cite{petry2021}

	\item global optimization of atomic clusters, including electronic spin multiplicities, via active learning and Gaussian process regression \cite{lourenco2021q,lourenco2021bq}

	\item derivative-free structural relaxation of water and small unbranched alkanes via kernel ridge regression and simulated annealing \cite{iype2019q}

	\item identification of low-energy point defects in solids via evolutionary algorithms, clustering, and Gaussian process regression \cite{arrigoni2021q}

	\item Monte Carlo simulations of the thermodynamics of thiolate-protected gold nanoclusters via minimal learning machine regression \cite{phlnmkh2020q}

	\item developing data-efficient machine-learning potentials at the example of Cs$^+$ in water via active learning and Gaussian process regression. \cite{zhai2020q} 

\end{itemize}

\section{Discussion and outlook}

\noindent
MBTR is a general representation (numerical description, feature set) of atomistic systems for fast accurate interpolation between quantum-mechanical calculations via ML.
It is based on distributions of $k$-atom terms stratified by chemical elements.
Despite, or because of, this simple principle, it is connected to many other representations, including 
CM~\cite{rtml2012}, 
BoB~\cite{hbrplmt2015},
histograms of distances, angles and dihedral angles~\cite{fhhgsdvkrv2017}, 
atom-centered symmetry functions~\cite{bp2007}, 
partial radial distribution functions~\cite{sgbsmg2014},
Faber-Christensen-Huang-von Lilienfeld representation~\cite{faber2018,christensen2020},
as well as cluster expansion~\cite{sdg1984}.
See Ref.~\citenum{lgr2021} for further details on these and other relationships.


MBTR represents whole molecules and crystals.
With increasing number of atoms, and thus degrees of freedom, this approach is likely to degrade, and exploitation of locality via prediction of additive atomic energy contributions becomes appealing.~\cite{bpkc2010,b2011d}
This requires representing local chemical environments~\cite{bkc2013}, for which MBTR can be modified \cite{jager2018,hxh2020,hjmfrgrf2020}.

We note in passing that problems in the training of ML models, such as outliers, can often be traced back to problems in the underlying reference calculations, such as unconverged fast Fourier transform grids or inconsistent settings (violating the assumption that a single function is being fitted), a phenomenon also observed by others.~\cite{perscom} 
This suggests that automated identification of errors in big datasets of electronic structure calculations via parametrization of ML models might be a general approach for validation of such datasets.
We rationalize this hypothesis by ML models identifying regularity (correlations) in data, and faulty calculations deviating in some way from correct ones.

Continuing advances in electronic structure codes and increasing availability of large-scale computing resources have led to large collections of {ab initio} calculations, such as 
Materials Project \cite{johcrdcgscp2013}, 
AFLOWlib~\cite{cshjctwxylmsdm2012}, 
Open Quantum Materials Database \cite{ksmtdarw2015}, 
and Novel Materials Discovery Laboratory \cite{ds2019}.
Representations such as MBTR are key to combine quantum mechanics with machine learning (QM/ML) for fast, accurate and precise interpolation in these settings.

\medskip

\begin{acknowledgments}
\noindent\textbf{Acknowledgments}
MR and HH thank Matthias Scheffler for helpful discussions and support.
MR thanks the Institute for Pure and Applied Mathematics (IPAM) for hospitality and support, participants of its program on Understanding Many-Particle Systems with Machine Learning for feedback, and Albert Bart{\'o}k-P{\'a}rtay, G{\'a}bor Cs{\'a}nyi, Alexander Shapeev, Alexandre Tkatchenko and the late Alessandro de Vita for extended discussions.
MR acknowledges funding from EU Horizon 2020 program grant 676580, The Novel Materials Discovery (NOMAD) Laboratory, a European Center of Excellence.
\end{acknowledgments}

\smallskip

\noindent
This article is an extended and updated version of the original arXiv preprint 1704.06439,
whose formal publication was delayed for non-technical reasons.

\medskip

\noindent\textbf{Data and Code Availability}
All datasets used in this study are publicly available.
Implementations of MBTR are available as part of the DScribe \cite{hjmfrgrf2020} and qmmlpack \cite{qmmlpackimpl} libraries.
Code to reproduce results of reported experiments is available at \href{https://github.com/hhaoyan/mbtr}{github.com/hhaoyan/mbtr}.

\clearpage


\begin{thebibliography}{92}

\bibitem[Ghahramani(2015)]{g2015}
Zoubin Ghahramani.
\newblock Probabilistic machine learning and artificial intelligence.
\newblock \emph{Nature}, 521\penalty0 (7553):\penalty0 452--459, 2015.
\newblock \doi{10.1038/nature14541}.

\bibitem[Jordan and Mitchell(2015)]{jm2015}
Michael~I. Jordan and Tom~M. Mitchell.
\newblock Machine learning: Trends, perspectives, and prospects.
\newblock \emph{Science}, 349\penalty0 (6245):\penalty0 255--260, 2015.
\newblock \doi{10.1126/science.aaa8415}.

\bibitem[Jinnouchi et~al.(2019)Jinnouchi, Karsai, and Kresse]{jkk2019q}
Ryosuke Jinnouchi, Ferenc Karsai, and Georg Kresse.
\newblock On-the-fly machine learning force field generation: Application to
  melting points.
\newblock \emph{Phys\BAP{}Rev\BAP{}B\BANE{}}, 100\penalty0 (1):\penalty0
  014105, 2019.
\newblock \doi{10.1103/physrevb.100.014105}.

\bibitem[Sendek et~al.(2019)Sendek, Cubuk, Antoniuk, Cheon, Cui, and
  Reed]{scaccr2019q}
Austin~D. Sendek, Ekin~D. Cubuk, Evan~R. Antoniuk, Gowoon Cheon, Yi~Cui, and
  Evan~J. Reed.
\newblock Machine learning-assisted discovery of solid {Li}-ion conducting
  materials.
\newblock \emph{Chem\BAP{}Mater\BAPE{}}, 31\penalty0 (2):\penalty0 342--352,
  2019.
\newblock \doi{10.1021/acs.chemmater.8b03272}.

\bibitem[Ramakrishnan et~al.(2015)Ramakrishnan, Dral, Rupp, and von
  Lilienfeld]{rdrl2015}
Raghunathan Ramakrishnan, Pavlo~O. Dral, Matthias Rupp, and O.~Anatole von
  Lilienfeld.
\newblock Big data meets quantum chemistry approximations: The
  {$\Delta$}-machine learning approach.
\newblock \emph{J\BAP{}Chem\BAP{}Theor\BAP{}Comput\BAPE{}}, 11\penalty0
  (5):\penalty0 2087--2096, 2015.
\newblock \doi{10.1021/acs.jctc.5b00099}.

\bibitem[Rupp et~al.(2012)Rupp, Tkatchenko, M{\"u}ller, and von
  Lilienfeld]{rtml2012}
Matthias Rupp, Alexandre Tkatchenko, Klaus-Robert M{\"u}ller, and O.~Anatole
  von Lilienfeld.
\newblock Fast and accurate modeling of molecular atomization energies with
  machine learning.
\newblock \emph{Phys\BAP{}Rev\BAP{}Lett\BAPE{}}, 108\penalty0 (5):\penalty0
  058301, 2012.
\newblock \doi{10.1103/PhysRevLett.108.058301}.

\bibitem[Rupp(2015{\natexlab{a}})]{r2015}
Matthias Rupp.
\newblock Machine learning for quantum mechanics in a nutshell.
\newblock \emph{Int\BAP{}J\BAP{}Quant\BAP{}Chem\BAPE{}}, 115\penalty0
  (16):\penalty0 1058--1073, 2015{\natexlab{a}}.
\newblock \doi{10.1002/qua.24954}.

\bibitem[Ramakrishnan et~al.(2014)Ramakrishnan, Dral, Rupp, and von
  Lilienfeld]{rdrl2014}
Raghunathan Ramakrishnan, Pavlo~O. Dral, Matthias Rupp, and O.~Anatole von
  Lilienfeld.
\newblock Quantum chemistry structures and properties of 134 kilo molecules.
\newblock \emph{Sci\BAP{}Data\BANE{}}, 1:\penalty0 140022, 2014.
\newblock \doi{10.1038/sdata.2014.22}.

\bibitem[Bart{\'o}k et~al.(2010)Bart{\'o}k, Payne, Kondor, and
  Cs{\'a}nyi]{bpkc2010}
Albert~P. Bart{\'o}k, Mike~C. Payne, Risi Kondor, and G{\'a}bor Cs{\'a}nyi.
\newblock {G}aussian approximation potentials: The accuracy of quantum
  mechanics, without the electrons.
\newblock \emph{Phys\BAP{}Rev\BAP{}Lett\BAPE{}}, 104\penalty0 (13):\penalty0
  136403, 2010.
\newblock \doi{10.1103/PhysRevLett.104.136403}.

\bibitem[Rupp et~al.(2015)Rupp, Ramakrishnan, and von Lilienfeld]{rrl2015}
Matthias Rupp, Raghunathan Ramakrishnan, and O.~Anatole von Lilienfeld.
\newblock Machine learning for quantum mechanical properties of atoms in
  molecules.
\newblock \emph{J\BAP{}Phys\BAP{}Chem\BAP{}Lett\BAPE{}}, 6\penalty0
  (16):\penalty0 3309--3313, 2015.
\newblock \doi{10.1021/acs.jpclett.5b01456}.

\bibitem[Bart{\'o}k et~al.(2013{\natexlab{a}})Bart{\'o}k, Gillan, Manby, and
  Cs{\'a}nyi]{bgmc2013}
Albert~P. Bart{\'o}k, Michael~J. Gillan, Frederick~R. Manby, and G{\'a}bor
  Cs{\'a}nyi.
\newblock Machine-learning approach for one- and two-body corrections to
  density functional theory: Applications to molecular and condensed water.
\newblock \emph{Phys\BAP{}Rev\BAP{}B\BANE{}}, 88\penalty0 (5):\penalty0 054104,
  2013{\natexlab{a}}.
\newblock \doi{10.1103/PhysRevB.88.054104}.

\bibitem[Rupp(2015{\natexlab{b}})]{r2015fq}
Matthias Rupp.
\newblock Machine learning for quantum mechanics in a nutshell.
\newblock \emph{Int\BAP{}J\BAP{}Quant\BAP{}Chem\BAPE{}}, 115\penalty0
  (16):\penalty0 1058--1073, 2015{\natexlab{b}}.
\newblock \doi{10.1002/qua.24954}.

\bibitem[Chmiela et~al.(2017)Chmiela, Tkatchenko, Sauceda, Poltavsky,
  Sch{\"u}tt, and M{\"u}ller]{ctspsm2017}
Stefan Chmiela, Alexandre Tkatchenko, Huziel~E. Sauceda, Igor Poltavsky,
  Kristof Sch{\"u}tt, and Klaus-Robert M{\"u}ller.
\newblock Machine learning of accurate energy-conserving molecular force
  fields.
\newblock \emph{Sci\BAP{}Adv\BAPE{}}, 3\penalty0 (5):\penalty0 e1603015, 2017.
\newblock \doi{10.1126/sciadv.1603015}.

\bibitem[Deringer et~al.(2021)Deringer, Bart{\'o}k, Bernstein, Wilkins,
  Ceriotti, and Cs{\'a}nyi]{deringer2021q}
Volker~L. Deringer, Albert~P. Bart{\'o}k, Noam Bernstein, David~M. Wilkins,
  Michele Ceriotti, and G{\'a}bor Cs{\'a}nyi.
\newblock {G}aussian process regression for materials and molecules.
\newblock \emph{Chem\BAP{}Rev\BAPE{}}, 121\penalty0 (16):\penalty0
  10073--10141, 2021.
\newblock \doi{10.1021/acs.chemrev.1c00022}.

\bibitem[Unke et~al.(2021)Unke, Chmiela, Sauceda, Gastegger, Poltavsky,
  Sch{\"u}tt, Tkatchenko, and M{\"u}ller]{unke2021q}
Oliver~T. Unke, Stefan Chmiela, Huziel~E. Sauceda, Michael Gastegger, Igor
  Poltavsky, Kristof~T. Sch{\"u}tt, Alexandre Tkatchenko, and Klaus-Robert
  M{\"u}ller.
\newblock Machine learning force fields.
\newblock \emph{Chem\BAP{}Rev\BAPE{}}, 121\penalty0 (16):\penalty0
  10142--10186, 2021.
\newblock \doi{10.1021/acs.chemrev.0c01111}.

\bibitem[Langer et~al.(2022)Langer, Goe{\ss}mann, and Rupp]{lgr2021}
Marcel~F. Langer, Alex Goe{\ss}mann, and Matthias Rupp.
\newblock Representations of molecules and materials for interpolation of
  quantum-mechanical simulations via machine learning.
\newblock \emph{npj\BAN{}Comput\BAP{}Mater\BAPE{}}, 8:\penalty0 41, 2022.
\newblock \doi{10.1038/s41524-022-00721-x}.

\bibitem[ker()]{kernelspace}
Kernel methods use a positive definite function (kernel) to implicitly define
  the Hilbert space. We focus on explicit numerical representations as input
  for vector kernels.

\bibitem[Moussa(2012)]{m2012e}
Jonathan~E. Moussa.
\newblock Comment on ``{F}ast and accurate modeling of molecular atomization
  energies with machine learning''.
\newblock \emph{Phys\BAP{}Rev\BAP{}Lett\BAPE{}}, 109\penalty0 (5):\penalty0
  059801, 2012.
\newblock \doi{10.1103/PhysRevLett.109.059801}.

\bibitem[Collins et~al.(2018)Collins, Gordon, von Lilienfeld, and
  Yaron]{cgvy2017}
Christopher~R. Collins, Geoffrey~J. Gordon, O.~Anatole von Lilienfeld, and
  David~J. Yaron.
\newblock Constant size descriptors for accurate machine learning models of
  molecular properties.
\newblock \emph{J\BAP{}Chem\BAP{}Phys\BAPE{}}, 148\penalty0 (24):\penalty0
  241718, 2018.
\newblock \doi{10.1063/1.5020441}.

\bibitem[Bart{\'o}k et~al.(2013{\natexlab{b}})Bart{\'o}k, Kondor, and
  Cs{\'a}nyi]{bkc2013}
Albert~P. Bart{\'o}k, Risi Kondor, and G{\'a}bor Cs{\'a}nyi.
\newblock On representing chemical environments.
\newblock \emph{Phys\BAP{}Rev\BAP{}B\BANE{}}, 87\penalty0 (18):\penalty0
  184115, 2013{\natexlab{b}}.
\newblock \doi{10.1103/PhysRevB.87.184115}.

\bibitem[von Lilienfeld et~al.(2015)von Lilienfeld, Ramakrishnan, Rupp, and
  Knoll]{lrrk2015}
O.~Anatole von Lilienfeld, Raghunathan Ramakrishnan, Matthias Rupp, and Aaron
  Knoll.
\newblock {F}ourier series of atomic radial distribution functions: A molecular
  fingerprint for machine learning models of quantum chemical properties.
\newblock \emph{Int\BAP{}J\BAP{}Quant\BAP{}Chem\BAPE{}}, 115\penalty0
  (16):\penalty0 1084--1093, 2015.
\newblock \doi{10.1002/qua.24912}.

\bibitem[Onat et~al.(2020)Onat, Ortner, and Kermode]{ook2020q}
Berk Onat, Christoph Ortner, and James~R. Kermode.
\newblock Sensitivity and dimensionality of atomic environment representations
  used for machine learning interatomic potentials.
\newblock \emph{J\BAP{}Chem\BAP{}Phys\BAPE{}}, 153\penalty0 (14):\penalty0
  144106, 2020.
\newblock \doi{10.1063/5.0016005}.

\bibitem[Hansen et~al.(2015)Hansen, Biegler, Ramakrishnan, Pronobis, von
  Lilienfeld, M{\"u}ller, and Tkatchenko]{hbrplmt2015}
Katja Hansen, Franziska Biegler, Raghunathan Ramakrishnan, Wiktor Pronobis,
  O.~Anatole von Lilienfeld, Klaus-Robert M{\"u}ller, and Alexandre Tkatchenko.
\newblock Machine learning predictions of molecular properties: Accurate
  many-body potentials and nonlocality in chemical space.
\newblock \emph{J\BAP{}Phys\BAP{}Chem\BAP{}Lett\BAPE{}}, 6:\penalty0
  2326--2331, 2015.
\newblock \doi{10.1021/acs.jpclett.5b00831}.

\bibitem[Todeschini and Consonni(2009)]{tc2009}
Roberto Todeschini and Viviana Consonni.
\newblock \emph{Handbook of Molecular Descriptors}.
\newblock Wiley, Weinheim, Germany, {2nd} edition, 2009.

\bibitem[Behler and Parrinello(2007)]{bp2007}
J{\"o}rg Behler and Michele Parrinello.
\newblock Generalized neural-network representation of high-dimensional
  potential-energy surfaces.
\newblock \emph{Phys\BAP{}Rev\BAP{}Lett\BAPE{}}, 98\penalty0 (14):\penalty0
  146401, 2007.
\newblock \doi{10.1103/PhysRevLett.98.146401}.

\bibitem[Faber et~al.(2017)Faber, Hutchison, Huang, Gilmer, Schoenholz, Dahl,
  Vinyals, Kearnes, Riley, and von Lilienfeld]{fhhgsdvkrv2017}
Felix~A. Faber, Luke Hutchison, Bing Huang, Justin Gilmer, Samuel~S.
  Schoenholz, George~E. Dahl, Oriol Vinyals, Steven Kearnes, Patrick~F. Riley,
  and O.~Anatole von Lilienfeld.
\newblock Prediction errors of molecular machine learning models lower than
  hybrid {DFT} error.
\newblock \emph{J\BAP{}Chem\BAP{}Theor\BAP{}Comput\BAPE{}}, 13\penalty0
  (11):\penalty0 5255--5264, 2017.
\newblock \doi{10.1021/acs.jctc.7b00577}.

\bibitem[Lumiaro et~al.(2021)Lumiaro, Todorovi{\'c}, Kurten, Vehkam{\"a}ki, and
  Rinke]{lumiaro2021q}
Emma Lumiaro, Milica Todorovi{\'c}, Theo Kurten, Hanna Vehkam{\"a}ki, and
  Patrick Rinke.
\newblock Predicting gas-particle partitioning coefficients of atmospheric
  molecules with machine learning.
\newblock \emph{Atmos\BAP{}Chem\BAP{}Phys\BAPE{}}, 21\penalty0 (17):\penalty0
  13227--13246, 2021.
\newblock \doi{10.5194/acp-21-13227-2021}.

\bibitem[Bahlke et~al.(2020)Bahlke, Mogos, Proppe, and Herrmann]{bahlke2020q}
Marc~Philipp Bahlke, Natnael Mogos, Jonny Proppe, and Carmen Herrmann.
\newblock Exchange spin coupling from {G}aussian process regression.
\newblock \emph{J\BAP{}Phys\BAP{}Chem\BAP{}A\BANE{}}, 124\penalty0
  (42):\penalty0 8708--8723, 2020.
\newblock \doi{10.1021/acs.jpca.0c05983}.

\bibitem[Petry et~al.(2021)Petry, Focassio, Schleder, Martinez, and
  Fazzio]{petry2021}
Romana Petry, Bruno Focassio, Gabriel~R. Schleder, Diego St{\'e}fani~T.
  Martinez, and Adalberto Fazzio.
\newblock Conformational analysis of tannic acid: Environment effects in
  electronic and reactivity properties.
\newblock \emph{J\BAP{}Chem\BAP{}Phys\BAPE{}}, 154\penalty0 (22):\penalty0
  224102, 2021.
\newblock \doi{10.1063/5.0045968}.

\bibitem[Louren{\c{c}}o et~al.(2021{\natexlab{a}})Louren{\c{c}}o, Herrera,
  Hosta{\v{s}}, Calaminici, K{\"o}ster, Tchagang, and Salahub]{lourenco2021q}
Maicon~Pierre Louren{\c{c}}o, Lizandra~Barrios Herrera, Ji{\v{r}}{\'{\i}}
  Hosta{\v{s}}, Patrizia Calaminici, Andreas~M. K{\"o}ster, Alain Tchagang, and
  Dennis~R. Salahub.
\newblock Taking the multiplicity inside the loop: active learning for
  structural and spin multiplicity elucidation of atomic clusters.
\newblock \emph{Theoretical Chemistry Accounts}, 140\penalty0 (8):\penalty0
  116, 2021{\natexlab{a}}.
\newblock \doi{10.1007/s00214-021-02820-2}.

\bibitem[Louren{\c{c}}o et~al.(2021{\natexlab{b}})Louren{\c{c}}o, Galv{\~{a}}o,
  Herrera, Hosta{\v{s}}, Tchagang, Silva, and Salahub]{lourenco2021bq}
Maicon~Pierre Louren{\c{c}}o, Breno R.~L. Galv{\~{a}}o, Lizandra~Barrios
  Herrera, Ji{\v{r}}{\'{\i}} Hosta{\v{s}}, Alain Tchagang, Mateus~X. Silva, and
  Dennis~R. Salahub.
\newblock A new active learning approach for global optimization of atomic
  clusters.
\newblock \emph{Theoretical Chemistry Accounts}, 140\penalty0 (6):\penalty0 62,
  2021{\natexlab{b}}.
\newblock \doi{10.1007/s00214-021-02766-5}.

\bibitem[Iype and Urolagin(2019)]{iype2019q}
Eldhose Iype and Siddhaling Urolagin.
\newblock Machine learning model for non-equilibrium structures and energies of
  simple molecules.
\newblock \emph{J\BAP{}Chem\BAP{}Phys\BAPE{}}, 150\penalty0 (2):\penalty0
  024307, 2019.
\newblock \doi{10.1063/1.5054968}.

\bibitem[Zhai et~al.(2020)Zhai, Caruso, Gao, and Paesani]{zhai2020q}
Yaoguang Zhai, Alessandro Caruso, Sicun Gao, and Francesco Paesani.
\newblock Active learning of many-body configuration space: Application to the
  {Cs$^+$}-water {MB-nrg} potential energy function as a case study.
\newblock \emph{J\BAP{}Chem\BAP{}Phys\BAPE{}}, 152\penalty0 (14):\penalty0
  144103, 2020.
\newblock \doi{10.1063/5.0002162}.

\bibitem[Honrao et~al.(2020)Honrao, Xie, and Hennig]{hxh2020}
Shreyas~J. Honrao, Stephen~R. Xie, and Richard~G. Hennig.
\newblock Augmenting machine learning of energy landscapes with local
  structural information.
\newblock \emph{J\BAP{}Appl\BAP{}Phys\BAPE{}}, 128\penalty0 (8):\penalty0
  085101, 2020.
\newblock \doi{10.1063/5.0012407}.

\bibitem[Mayr and Gagliardi(2021)]{mayr2021q}
Felix Mayr and Alessio Gagliardi.
\newblock Global property prediction: A benchmark study on open-source,
  perovskite-like datasets.
\newblock \emph{ACS\BAN{}Omega\BANE{}}, 6\penalty0 (19):\penalty0 12722--12732,
  2021.
\newblock \doi{10.1021/acsomega.1c00991}.

\bibitem[Arrigoni and Madsen(2021)]{arrigoni2021q}
Marco Arrigoni and Georg K.~H. Madsen.
\newblock Evolutionary computing and machine learning for discovering of
  low-energy defect configurations.
\newblock \emph{npj\BAN{}Comput\BAP{}Mater\BAPE{}}, 7\penalty0 (1):\penalty0
  71, 2021.
\newblock \doi{10.1038/s41524-021-00537-1}.

\bibitem[Pihlajam{\"a}ki et~al.(2020)Pihlajam{\"a}ki, H{\"a}m{\"a}l{\"a}inen,
  Linja, Nieminen, Malola, K{\"a}rkk{\"a}inen, and H{\"a}kkinen]{phlnmkh2020q}
Antti Pihlajam{\"a}ki, Joonas H{\"a}m{\"a}l{\"a}inen, Joakim Linja, Paavo
  Nieminen, Sami Malola, Tommi K{\"a}rkk{\"a}inen, and Hannu H{\"a}kkinen.
\newblock {M}onte {C}arlo simulations of {Au$_{38}$({SCH}$_3$)$_{24}$}
  nanocluster using distance-based machine learning methods.
\newblock \emph{J\BAP{}Phys\BAP{}Chem\BAP{}A\BANE{}}, 124\penalty0
  (23):\penalty0 4827--4836, 2020.
\newblock \doi{10.1021/acs.jpca.0c01512}.

\bibitem[Montavon et~al.(2013)Montavon, Rupp, Gobre, Vazquez-Mayagoitia,
  Hansen, Tkatchenko, M{\"u}ller, and von Lilienfeld]{mrgvmhtmvl2013}
Gr{\'e}goire Montavon, Matthias Rupp, Vivekanand Gobre, Alvaro
  Vazquez-Mayagoitia, Katja Hansen, Alexandre Tkatchenko, Klaus-Robert
  M{\"u}ller, and O.~Anatole von Lilienfeld.
\newblock Machine learning of molecular electronic properties in chemical
  compound space.
\newblock \emph{New\BAN{}J\BAP{}Phys\BAPE{}}, 15\penalty0 (9):\penalty0 095003,
  2013.
\newblock \doi{10.1088/1367-2630/15/9/095003}.

\bibitem[Ghiringhelli et~al.(2015)Ghiringhelli, Vybiral, Levchenko, Draxl, and
  Scheffler]{gvlds2015}
Luca~M. Ghiringhelli, Jan Vybiral, Sergey~V. Levchenko, Claudia Draxl, and
  Matthias Scheffler.
\newblock Big data of materials science: Critical role of the descriptor.
\newblock \emph{Phys\BAP{}Rev\BAP{}Lett\BAPE{}}, 114\penalty0 (10):\penalty0
  105503, 2015.
\newblock \doi{10.1103/PhysRevLett.114.105503}.

\bibitem[Pozdnyakov et~al.(2020)Pozdnyakov, Willatt, Bart{\'{o}}k, Ortner,
  Cs{\'{a}}nyi, and Ceriotti]{pwbocc2020q}
Sergey~N. Pozdnyakov, Michael~J. Willatt, Albert~P. Bart{\'{o}}k, Christoph
  Ortner, G{\'{a}}bor Cs{\'{a}}nyi, and Michele Ceriotti.
\newblock Incompleteness of atomic structure representations.
\newblock \emph{Phys\BAP{}Rev\BAP{}Lett\BAPE{}}, 125\penalty0 (16):\penalty0
  166001, 2020.
\newblock \doi{10.1103/physrevlett.125.166001}.

\bibitem[Cubuk et~al.(2015)Cubuk, Schoenholz, Rieser, Malone, Rottler, Durian,
  Kaxiras, and Liu]{csrmrdkl2015}
Ekin~D. Cubuk, Samuel~S. Schoenholz, Jennifer~M. Rieser, Brad~D. Malone, Joerg
  Rottler, Douglas~J. Durian, Efthimios Kaxiras, and Andrea~J. Liu.
\newblock Identifying structural flow defects in disordered solids using
  machine-learning methods.
\newblock \emph{Phys\BAP{}Rev\BAP{}Lett\BAPE{}}, 114\penalty0 (10):\penalty0
  108001, 2015.
\newblock \doi{10.1103/PhysRevLett.114.108001}.

\bibitem[Huang and von Lilienfeld(2016)]{hl2016}
Bing Huang and O.~Anatole von Lilienfeld.
\newblock Communication: Understanding molecular representations in machine
  learning: the role of uniqueness and target similarity.
\newblock \emph{J\BAP{}Chem\BAP{}Phys\BAPE{}}, 145\penalty0 (16):\penalty0
  161102, 2016.
\newblock \doi{10.1063/1.4964627}.

\bibitem[Yao et~al.(2017)Yao, Herr, and Parkhill]{yhp2017}
Kun Yao, John~E. Herr, and John Parkhill.
\newblock The many-body expansion combined with neural networks.
\newblock \emph{J\BAP{}Chem\BAP{}Phys\BAPE{}}, 146\penalty0 (1):\penalty0
  014106, 2017.
\newblock \doi{10.1063/1.4973380}.

\bibitem[sca()]{scalargeomf}
We use scalar geometry functions~$g_k$ for convenience; assigning vectors would
  simply increase the rank of the tensor. The product structure $w_k(\vec{i})
  \mathcal{D}(x,g_k(\vec{i}))$ allows efficient implementation as $\mathcal{D}$
  does not depend on~$\mathcal{M}$.

\bibitem[Faber et~al.(2018)Faber, Christensen, Huang, and von
  Lilienfeld]{faber2018}
Felix~A. Faber, Anders~S. Christensen, Bing Huang, and O.~Anatole von
  Lilienfeld.
\newblock Alchemical and structural distribution based representation for
  universal quantum machine learning.
\newblock \emph{J\BAP{}Chem\BAP{}Phys\BAPE{}}, 148\penalty0 (24):\penalty0
  241717, 2018.
\newblock \doi{10.1063/1.5020710}.

\bibitem[Herr et~al.(2019)Herr, Koh, Yao, and Parkhill]{hkyp2019q}
John~E. Herr, Kevin Koh, Kun Yao, and John Parkhill.
\newblock Compressing physics with an autoencoder: Creating an atomic species
  representation to improve machine learning models in the chemical sciences.
\newblock \emph{J\BAP{}Chem\BAP{}Phys\BAPE{}}, 151\penalty0 (6--7):\penalty0
  455--472, 2019.
\newblock \doi{10.1063/1.5108803}.

\bibitem[Christensen et~al.(2020)Christensen, Bratholm, Faber, and von
  Lilienfeld]{christensen2020}
Anders~S. Christensen, Lars~A. Bratholm, Felix~A. Faber, and O.~Anatole von
  Lilienfeld.
\newblock {FCHL} revisited: Faster and more accurate quantum machine learning.
\newblock \emph{J\BAP{}Chem\BAP{}Phys\BAPE{}}, 152\penalty0 (4):\penalty0
  044107, 2020.
\newblock \doi{10.1063/1.5126701}.

\bibitem[ind()]{indexincell}
Effectively representing one unit cell, including influence of surrounding
  cells on it, in accordance with computed properties being reported per cell.

\bibitem[exp()]{expweighting}
Exponential weighting was motivated by the exponential decay of screened
  Coulombic interactions in solids.

\bibitem[Frostig et~al.(2018)Frostig, Johnson, and Leary]{jax}
Roy Frostig, Matthew~James Johnson, and Chris Leary.
\newblock Compiling machine learning programs via high-level tracing.
\newblock In \emph{1st Conference on Systems and Machine Learning (SysML~2018),
  Stanford, California, February 15--16}, 2018.
\newblock URL \url{https://mlsys.org/Conferences/doc/2018/146.pdf}.

\bibitem[Paszke et~al.(2019)Paszke, Gross, Massa, Lerer, Bradbury, Chanan,
  Killeen, Lin, Gimelshein, Antiga, Desmaison, Kopf, Yang, DeVito, Raison,
  Tejani, Chilamkurthy, Steiner, Fang, Bai, and Chintala]{pytorch}
Adam Paszke, Sam Gross, Francisco Massa, Adam Lerer, James Bradbury, Gregory
  Chanan, Trevor Killeen, Zeming Lin, Natalia Gimelshein, Luca Antiga, Alban
  Desmaison, Andreas Kopf, Edward Yang, Zachary DeVito, Martin Raison, Alykhan
  Tejani, Sasank Chilamkurthy, Benoit Steiner, Lu~Fang, Junjie Bai, and Soumith
  Chintala.
\newblock {PyTorch}: An imperative style, high-performance deep learning
  library.
\newblock In Hanna Wallach, Hugo Larochelle, Alina Beygelzimer, Florence
  d'Alch{\'{e}} Buc, Emily Fox, and Roman Garnett, editors, \emph{Advances in
  Neural Information Processing Systems~32 (NeurIPS~2019), Montr{\'e}al,
  Canada, December~8--14}, pages 8024--8035. Curran Associates, 2019.
\newblock URL \url{https://neurips.cc/Conferences/2019}.

\bibitem[Perdew et~al.(1996{\natexlab{a}})Perdew, Burke, and
  Ernzerhof]{pbe1996}
John~P. Perdew, Kieron Burke, and Matthias Ernzerhof.
\newblock Generalized gradient approximation made simple.
\newblock \emph{Phys\BAP{}Rev\BAP{}Lett\BAPE{}}, 77\penalty0 (18):\penalty0
  3865--3868, 1996{\natexlab{a}}.
\newblock \doi{10.1103/PhysRevLett.77.3865}.

\bibitem[Perdew et~al.(1996{\natexlab{b}})Perdew, Ernzerhof, and
  Burke]{peb1996}
John~P. Perdew, Matthias Ernzerhof, and Kieron Burke.
\newblock Rationale for mixing exact exchange with density functional
  approximations.
\newblock \emph{J\BAP{}Chem\BAP{}Phys\BAPE{}}, 105\penalty0 (22):\penalty0
  9982--9985, 1996{\natexlab{b}}.

\bibitem[Adamo and Barone(1999)]{ab1999}
Carlo Adamo and Vincenzo Barone.
\newblock Toward reliable density functional methods without adjustable
  parameters: The {PBE0} model.
\newblock \emph{J\BAP{}Chem\BAP{}Phys\BAPE{}}, 110\penalty0 (13):\penalty0
  6158--6170, 1999.
\newblock \doi{10.1063/1.478522}.

\bibitem[De et~al.(2016)De, Bart{\'o}k, Cs{\'a}nyi, and Ceriotti]{dbcc2016}
Sandip De, Albert~P. Bart{\'o}k, G{\'a}bor Cs{\'a}nyi, and Michele Ceriotti.
\newblock Comparing molecules and solids across structural and alchemical
  space.
\newblock \emph{Phys\BAP{}Chem\BAP{}Chem\BAP{}Phys\BAPE{}}, 18\penalty0
  (20):\penalty0 13754--13769, 2016.
\newblock \doi{10.1039/C6CP00415F}.

\bibitem[Faber et~al.(2016)Faber, Lindmaa, von Lilienfeld, and
  Armiento]{flva2016}
Felix~A. Faber, Alexander Lindmaa, O.~Anatole von Lilienfeld, and Rickard
  Armiento.
\newblock Machine learning energies of 2 million elpasolite ({ABC$_2$D$_6$})
  crystals.
\newblock \emph{Phys\BAP{}Rev\BAP{}Lett\BAPE{}}, 117\penalty0 (13):\penalty0
  135502, 2016.
\newblock \doi{10.1103/PhysRevLett.117.135502}.

\bibitem[elp()]{elpasolitedataset}
Dataset ABC2D6-16, available at \protect\url{http://qmml.org}.

\bibitem[Saal et~al.(2013)Saal, Kirklin, Aykol, Meredig, and
  Wolverton]{skamw2013}
James~E. Saal, Scott Kirklin, Muratahan Aykol, Bryce Meredig, and Chris
  Wolverton.
\newblock Materials design and discovery with high-throughput density
  functional theory: The open quantum materials database ({OQMD}).
\newblock \emph{J\BAP{}Miner\BAP{}Met\BAP{}Mater\BAP{}Soc\BAPE{}}, 65\penalty0
  (11):\penalty0 1501--1509, 2013.
\newblock \doi{10.1007/s11837-013-0755-4}.

\bibitem[Kirklin et~al.(2015)Kirklin, Saal, Meredig, Thompson, Doak, Aykol,
  R{\"u}hl, and Wolverton]{ksmtdarw2015}
Scott Kirklin, James~E. Saal, Bryce Meredig, Alex Thompson, Jeff~W. Doak,
  Muratahan Aykol, Stephan R{\"u}hl, and Chris Wolverton.
\newblock The open quantum materials database ({OQMD}): assessing the accuracy
  of {DFT} formation energies.
\newblock \emph{npj\BAN{}Comput\BAP{}Mater\BAPE{}}, 1:\penalty0 15010, 2015.
\newblock \doi{10.1038/npjcompumats.2015.10}.

\bibitem[Sch{\"u}tt et~al.(2021)Sch{\"u}tt, Unke, and Gastegger]{sug2021q}
Kristof~T. Sch{\"u}tt, Oliver~T. Unke, and Michael Gastegger.
\newblock Equivariant message passing for the prediction of tensorial
  properties and molecular spectra.
\newblock In Marina Meila and Tong Zhang, editors, \emph{Proceedings of the
  38th International Conference on Machine Learning (ICML~2021), virtual,
  July~18--24}, pages 9377--9388. Proceedings of Machine Learning Research,
  2021.
\newblock URL \url{https://proceedings.mlr.press/v139/schutt21a.html}.

\bibitem[Chmiela et~al.(2018)Chmiela, Sauceda, M{\"u}ller, and
  Tkatchenko]{chmiela2018}
Stefan Chmiela, Huziel~E. Sauceda, Klaus-Robert M{\"u}ller, and Alexandre
  Tkatchenko.
\newblock Towards exact molecular dynamics simulations with machine-learned
  force fields.
\newblock \emph{Nat\BAP{}Commun\BAPE{}}, 9:\penalty0 3887, 2018.
\newblock \doi{10.1038/s41467-018-06169-2}.

\bibitem[Sch{\"u}tt et~al.(2017)Sch{\"u}tt, Arbabzadah, Chmiela, M{\"u}ller,
  and Tkatchenko]{sacmt2017}
Kristof~T. Sch{\"u}tt, Farhad Arbabzadah, Stefan Chmiela, Klaus~R. M{\"u}ller,
  and Alexandre Tkatchenko.
\newblock Quantum-chemical insights from deep tensor neural networks.
\newblock \emph{Nat\BAP{}Comm\BAPE{}}, 8:\penalty0 13890, 2017.
\newblock \doi{10.1038/ncomms13890}.

\bibitem[Tkatchenko and Scheffler(2009)]{ts2009}
Alexandre Tkatchenko and Matthias Scheffler.
\newblock Accurate molecular van der {W}aals interactions from ground-state
  electron density and free-atom reference data.
\newblock \emph{Phys\BAP{}Rev\BAP{}Lett\BAPE{}}, 102\penalty0 (7):\penalty0
  073005, 2009.
\newblock \doi{10.1103/PhysRevLett.102.073005}.

\bibitem[Snyder et~al.(2012)Snyder, Rupp, Hansen, M{\"u}ller, and
  Burke]{srhmb2012}
John~C. Snyder, Matthias Rupp, Katja Hansen, Klaus-Robert M{\"u}ller, and
  Kieron Burke.
\newblock Finding density functionals with machine learning.
\newblock \emph{Phys\BAP{}Rev\BAP{}Lett\BAPE{}}, 108\penalty0 (25):\penalty0
  253002, 2012.
\newblock \doi{10.1103/PhysRevLett.108.253002}.

\bibitem[Glielmo et~al.(2017)Glielmo, Sollich, and Vita]{gsd2017}
Aldo Glielmo, Peter Sollich, and Alessandro~De Vita.
\newblock Accurate interatomic force fields via machine learning with covariant
  kernels.
\newblock \emph{Phys\BAP{}Rev\BAP{}B\BANE{}}, 95\penalty0 (21):\penalty0
  214302, 2017.
\newblock \doi{10.1103/PhysRevB.95.214302}.

\bibitem[Hart et~al.(2013)Hart, Curtarolo, Massalski, and Levy]{hcml2013}
Gus~L.W. Hart, Stefano Curtarolo, Thaddeus~B. Massalski, and Ohad Levy.
\newblock Comprehensive search for new phases and compounds in binary alloy
  systems based on platinum-group metals, using a computational
  first-principles approach.
\newblock \emph{Phys\BAP{}Rev.~X\BANE{}}, 3:\penalty0 041035, 2013.
\newblock \doi{10.1103/PhysRevX.3.041035}.

\bibitem[Settles(2012)]{s2012b}
Burr Settles.
\newblock \emph{Active Learning}, volume~18 of \emph{Synthesis Lectures on
  Artificial Intelligence and Machine Learning}.
\newblock Morgan \& Claypool, 2012.
\newblock \doi{10.2200/S00429ED1V01Y201207AIM018}.

\bibitem[Ulissi et~al.(2016)Ulissi, Singh, Tsai, and N{\o}rskov]{ustk2016q}
Zachary~W. Ulissi, Aayush~R. Singh, Charlie Tsai, and Jens~K. N{\o}rskov.
\newblock Automated discovery and construction of surface phase diagrams using
  machine learning.
\newblock \emph{J\BAP{}Phys\BAP{}Chem\BAP{}Lett\BAPE{}}, 7\penalty0
  (19):\penalty0 3931--3935, 2016.
\newblock \doi{10.1021/acs.jpclett.6b01254}.

\bibitem[Kolsbjerg et~al.(2018)Kolsbjerg, Peterson, and Hammer]{kph2018q}
Esben~L. Kolsbjerg, Andrew~A. Peterson, and Bj{\o}rk Hammer.
\newblock Neural-network-enhanced evolutionary algorithm applied to supported
  metal nanoparticles.
\newblock \emph{Phys\BAP{}Rev\BAP{}B\BANE{}}, 97\penalty0 (19):\penalty0
  195424, 2018.
\newblock \doi{10.1103/PhysRevB.97.195424}.

\bibitem[Denzel and K{\"a}stner(2018)]{dk2018q}
Alexander Denzel and Johannes K{\"a}stner.
\newblock {G}aussian process regression for geometry optimization.
\newblock \emph{J\BAP{}Chem\BAP{}Phys\BAPE{}}, 148\penalty0 (9):\penalty0
  094114, 2018.
\newblock \doi{10.1063/1.5017103}.

\bibitem[Schmitz and Christiansen(2018)]{sc2018q}
Gunnar Schmitz and Ove Christiansen.
\newblock {G}aussian process regression to accelerate geometry optimizations
  relying on numerical differentiation.
\newblock \emph{J\BAP{}Chem\BAP{}Phys\BAPE{}}, 148\penalty0 (24):\penalty0
  241704, 2018.
\newblock \doi{10.1063/1.5009347}.

\bibitem[Yoon and Ulissi(2020)]{yu2020q}
Junwoong Yoon and Zachary~W. Ulissi.
\newblock Differentiable optimization for the prediction of ground state
  structures ({DOGSS}).
\newblock \emph{Phys\BAP{}Rev\BAP{}Lett\BAPE{}}, 125\penalty0 (17):\penalty0
  173001, 2020.
\newblock \doi{10.1103/physrevlett.125.173001}.

\bibitem[Mortensen et~al.(2020)Mortensen, Meldgaard, Bisbo, Christiansen, and
  Hammer]{mmbch2020q}
Henrik~Lund Mortensen, S{\o}ren~Ager Meldgaard, Malthe~Kj{\ae}r Bisbo,
  Mads-Peter~V. Christiansen, and Bj{\o}rk Hammer.
\newblock Atomistic structure learning algorithm with surrogate energy model
  relaxation.
\newblock \emph{Phys\BAP{}Rev\BAP{}B\BANE{}}, 102\penalty0 (7):\penalty0
  075427, 2020.
\newblock \doi{10.1103/physrevb.102.075427}.

\bibitem[Huang et~al.(2022)Huang, Bao, and Tristan]{huang2022}
Daniel Huang, Junwei~Lucas Bao, and Jean-Baptiste Tristan.
\newblock Geometry meta-optimization.
\newblock \emph{J\BAP{}Chem\BAP{}Phys\BAPE{}}, 156\penalty0 (13):\penalty0
  134109, 2022.
\newblock \doi{10.1063/5.0087165}.

\bibitem[Hao et~al.(2022)Hao, He, Roitberg, Zhang, and Wang]{hao2022}
Dongxiao Hao, Xibing He, Adrian~E. Roitberg, Shengli Zhang, and Junmei Wang.
\newblock Development and evaluation of geometry optimization algorithms in
  conjunction with {ANI} potentials.
\newblock \emph{J\BAP{}Chem\BAP{}Theor\BAP{}Comput\BAPE{}}, 18\penalty0
  (2):\penalty0 978--991, 2022.
\newblock \doi{10.1021/acs.jctc.1c01043}.

\bibitem[Born and K{\"a}stner(2021)]{born2021q}
Daniel Born and Johannes K{\"a}stner.
\newblock Geometry optimization in internal coordinates based on {G}aussian
  process regression: Comparison of two approaches.
\newblock \emph{J\BAP{}Chem\BAP{}Theor\BAP{}Comput\BAPE{}}, 17\penalty0
  (9):\penalty0 5955--5967, 2021.
\newblock \doi{10.1021/acs.jctc.1c00517}.

\bibitem[Stuke et~al.(2020)Stuke, Kunkel, Golze, Todorovi{\'{c}}, Margraf,
  Reuter, Rinke, and Oberhofer]{skgtmrro2020q}
Annika Stuke, Christian Kunkel, Dorothea Golze, Milica Todorovi{\'{c}},
  Johannes~T. Margraf, Karsten Reuter, Patrick Rinke, and Harald Oberhofer.
\newblock Atomic structures and orbital energies of 61,489 crystal-forming
  organic molecules.
\newblock \emph{Sci\BAP{}Data\BANE{}}, 17\penalty0 (2--4):\penalty0 83--92,
  2020.
\newblock \doi{10.1038/s41597-020-0385-y}.

\bibitem[Rahaman and Gagliardi(2020)]{rahaman2020q}
Obaidur Rahaman and Alessio Gagliardi.
\newblock Deep learning total energies and orbital energies of large organic
  molecules using hybridization of molecular fingerprints.
\newblock \emph{J\BAP{}Chem\BAP{}Inf\BAP{}Model\BAPE{}}, 60\penalty0
  (12):\penalty0 5971--5983, 2020.
\newblock \doi{10.1021/acs.jcim.0c00687}.

\bibitem[Jung et~al.(2020)Jung, Stocker, Kunkel, Oberhofer, Han, Reuter, and
  Margraf]{jskohrm2020q}
Hyunwook Jung, Sina Stocker, Christian Kunkel, Harald Oberhofer, Byungchan Han,
  Karsten Reuter, and Johannes~T. Margraf.
\newblock Size-extensive molecular machine learning with global
  representations.
\newblock \emph{ChemSystemsChem\BANE{}}, 2\penalty0 (4):\penalty0 e1900052,
  2020.
\newblock \doi{10.1002/syst.201900052}.

\bibitem[Yaghoobi and Alaei(2022)]{yaghoobi2022}
Mostafa Yaghoobi and Mojtaba Alaei.
\newblock Machine learning for compositional disorder: A comparison between
  different descriptors and machine learning frameworks.
\newblock \emph{Comput\BAP{}Mater\BAP{}Sci\BAPE{}}, 207:\penalty0 111284, 2022.
\newblock \doi{10.1016/j.commatsci.2022.111284}.

\bibitem[Sch{\"u}tt et~al.(2014)Sch{\"u}tt, Glawe, Brockherde, Sanna,
  M{\"u}ller, and Gross]{sgbsmg2014}
Kristof~T. Sch{\"u}tt, Henning Glawe, Felix Brockherde, Antonio Sanna,
  Klaus-Robert M{\"u}ller, and Eberhard~K.U. Gross.
\newblock How to represent crystal structures for machine learning: Towards
  fast prediction of electronic properties.
\newblock \emph{Phys\BAP{}Rev\BAP{}B\BANE{}}, 89\penalty0 (20):\penalty0
  205118, 2014.
\newblock \doi{10.1103/PhysRevB.89.205118}.

\bibitem[Sanchez et~al.(1984)Sanchez, Ducastelle, and Gratias]{sdg1984}
Juan~M. Sanchez, Fran{\c{c}}ois Ducastelle, and Denis Gratias.
\newblock Generalized cluster description of multicomponent systems.
\newblock \emph{Phys\BAP{}Stat\BAP{}Mech\BAP{}Appl\BAPE{}}, 128\penalty0
  (1--2):\penalty0 334--350, 1984.
\newblock \doi{10.1016/0378-4371(84)90096-7}.

\bibitem[Behler(2011)]{b2011d}
J{\"o}rg Behler.
\newblock Neural network potential-energy surfaces in chemistry: a tool for
  large-scale simulations.
\newblock \emph{Phys\BAP{}Chem\BAP{}Chem\BAP{}Phys\BAPE{}}, 13\penalty0
  (40):\penalty0 17930--17955, 2011.
\newblock \doi{10.1039/C1CP21668F}.

\bibitem[J{\"a}ger et~al.(2018)J{\"a}ger, Morooka, Federici-Canova, Himanen,
  and Foster]{jager2018}
Marc O.~J. J{\"a}ger, Eiaki~V. Morooka, Filippo Federici-Canova, Lauri Himanen,
  and Adam~S. Foster.
\newblock Machine learning hydrogen adsorption on nanoclusters through
  structural descriptors.
\newblock \emph{npj\BAN{}Comput\BAP{}Mater\BAPE{}}, 4:\penalty0 37, 2018.
\newblock \doi{10.1038/s41524-018-0096-5}.

\bibitem[Himanen et~al.(2019)Himanen, J{\"{a}}ger, Morooka, Canova, Ranawat,
  Gao, Rinke, and Foster]{hjmfrgrf2020}
Lauri Himanen, Marc~O.J. J{\"{a}}ger, Eiaki~V. Morooka, Filippo~Federici
  Canova, Yashasvi~S. Ranawat, David~Z. Gao, Patrick Rinke, and Adam~S. Foster.
\newblock {DScribe}: Library of descriptors for machine learning in materials
  science.
\newblock \emph{Comput\BAP{}Phys\BAP{}Comm\BAPE{}}, 247:\penalty0 106949, 2019.
\newblock \doi{10.1016/j.cpc.2019.106949}.

\bibitem[per()]{perscom}
Independent personal communications by J{\"o}rg Behler, G{\'a}bor Cs{\'a}nyi,
  and Ekin Do{\v{g}}u\c{s} \c{C}ubuk.

\bibitem[Jain et~al.(2013)Jain, Ong, Hautier, Chen, Richards, Dacek, Cholia,
  Gunter, Skinner, Ceder, and Persson]{johcrdcgscp2013}
Anubhav Jain, Shyue~Ping Ong, Geoffroy Hautier, Wei Chen, William~Davidson
  Richards, Stephen Dacek, Shreyas Cholia, Dan Gunter, David Skinner, Gerbrand
  Ceder, and Kristin~A. Persson.
\newblock Commentary: The materials project: A materials genome approach to
  accelerating materials innovation.
\newblock \emph{{APL}\BAN{}Mater\BAPE{}}, 1:\penalty0 011002, 2013.
\newblock \doi{10.1063/1.4812323}.

\bibitem[Curtarolo et~al.(2012)Curtarolo, Setyawan, Hart, Jahnatek, Chepulskii,
  Taylor, Wang, Xue, Yang, Levy, Mehl, Stokes, Demchenko, and
  Morgan]{cshjctwxylmsdm2012}
Stefano Curtarolo, Wahyu Setyawan, Gus L.~W. Hart, Michal Jahnatek, Roman~V.
  Chepulskii, Richard~H. Taylor, Shidong Wang, Junkai Xue, Kesong Yang, Ohad
  Levy, Michael~J. Mehl, Harold~T. Stokes, Denis~O. Demchenko, and Dane Morgan.
\newblock {AFLOW}: An automatic framework for high-throughput materials
  discovery.
\newblock \emph{Comput\BAP{}Mater\BAP{}Sci\BAPE{}}, 58\penalty0 (6):\penalty0
  218--226, 2012.
\newblock \doi{10.1016/j.commatsci.2012.02.005}.

\bibitem[Draxl and Scheffler(2019)]{ds2019}
  Claudia Draxl and Matthias Scheffler.
  \newblock The {NOMAD} laboratory: from data sharing to artificial intelligence.
  \newblock \emph{J\BAP{}Phys\BAP{}Mater\BAPE{}}, 2(3):\penalty0 036001, 2019.
  \newblock \doi{10.1088/2515-7639/ab13bb}.

\bibitem[qmm()]{qmmlpackimpl}
Available as part of the software \texttt{qmmlpack} (quantum mechanics machine
  learning package) at \url{https://gitlab.com/qmml/qmmlpack} under the Apache
  2.0 license.

\end{thebibliography}

\providecommand{\BAN}{\,\discretionary{}{}{}}
  \providecommand{\BAP}{.\,\discretionary{}{}{}} \providecommand{\BANE}{}
  \providecommand{\BAPE}{.}

\providecommand{\natexlab}[1]{#1}
\providecommand{\url}[1]{\texttt{#1}}
\expandafter\ifx\csname urlstyle\endcsname\relax
  \providecommand{\doi}[1]{doi: #1}\else
  \providecommand{\doi}{doi: \begingroup \urlstyle{rm}\Url}\fi

\onecolumngrid

\clearpage
\pagestyle{empty}%
\includepdf[pages=1,pagecommand={}]{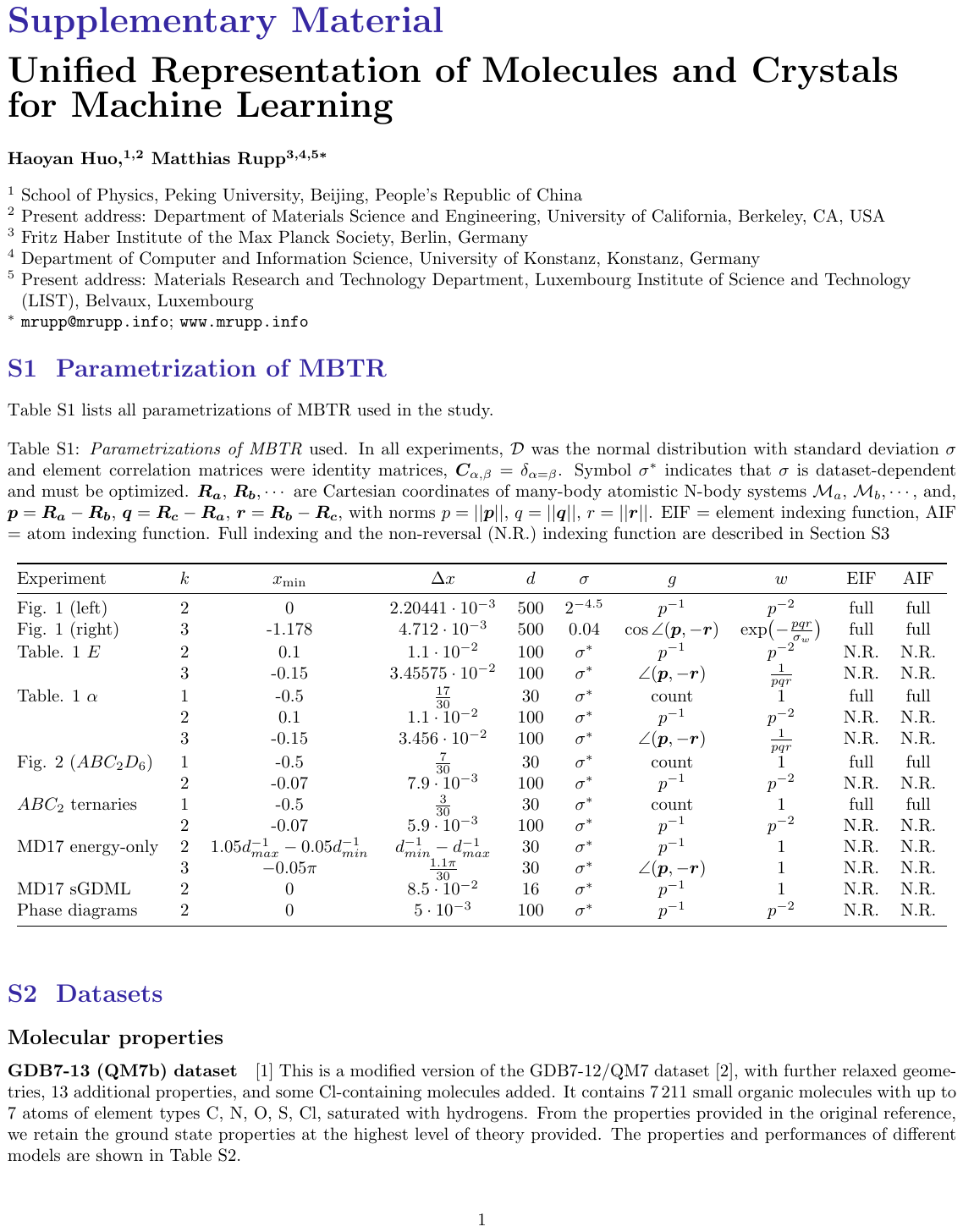}
\includepdf[pages=2,pagecommand={}]{supplement}
\includepdf[pages=3,pagecommand={}]{supplement}
\includepdf[pages=4,pagecommand={}]{supplement}
\includepdf[pages=5,pagecommand={}]{supplement}
\includepdf[pages=6,pagecommand={}]{supplement}

\end{document}